\documentclass[aps,twocolumn,pra,superscriptaddress,showpacs,tightenlines]{revtex4-1}
\usepackage{amsmath}
\usepackage{graphicx}
\usepackage{color}
\usepackage{amsfonts}
\usepackage{txfonts}
\usepackage[colorlinks,citecolor=blue]{hyperref}
\begin{document}

\title{Tunable optomechanically induced transparency by controlling the dark-mode effect}
\author{Deng-Gao Lai}
\affiliation{Key Laboratory of Low-Dimensional Quantum Structures and Quantum Control of Ministry of Education, Key Laboratory for Matter Microstructure and Function of Hunan Province, Department of Physics and Synergetic Innovation Center for Quantum Effects and Applications, Hunan Normal University, Changsha 410081, China}
\affiliation{Theoretical Quantum Physics Laboratory, RIKEN, Saitama 351-0198, Japan}

\author{Xin Wang}
\affiliation{Theoretical Quantum Physics Laboratory, RIKEN, Saitama 351-0198, Japan}
\affiliation{Institute of Quantum Optics and Quantum Information, School of Science, Xi An Jiaotong University, Xi An 710049, China}

\author{Wei Qin}
\affiliation{Theoretical Quantum Physics Laboratory, RIKEN, Saitama 351-0198, Japan}

\author{Bang-Pin Hou}
\email{bphou@sicnu.edu.cn}
\affiliation{College of Physics and Electronic Engineering, Institute of Solid State Physics, Sichuan Normal University, Chengdu 610068, P. R. China}

\author{Franco Nori}
\affiliation{Theoretical Quantum Physics Laboratory, RIKEN, Saitama 351-0198, Japan}
\affiliation{Physics Department, The University of Michigan, Ann Arbor, Michigan 48109-1040, USA}

\author{Jie-Qiao Liao}
\email{jqliao@hunnu.edu.cn}
\affiliation{Key Laboratory of Low-Dimensional Quantum Structures and Quantum Control of Ministry of Education, Key Laboratory for Matter Microstructure and Function of Hunan Province, Department of Physics and Synergetic Innovation Center for Quantum Effects and Applications, Hunan Normal University, Changsha 410081, China}

\begin{abstract}
We study tunable optomechanically-induced transparency by controlling the dark-mode effect induced by two mechanical modes coupled to a common cavity field. This is realized by introducing a phase-dependent phonon-exchange interaction, which is used to form a loop-coupled configuration. Combining this phase-dependent coupling with the optomechanical interactions, the dark-mode effect can be controlled by the quantum interference effect. In particular, the dark-mode effect in this two-mechanical-mode optomechanical system can lead to a double-amplified optomechanically induced transparency (OMIT) window and a higher efficiency of the second-order sideband in comparison with the standard optomechanical system. This is because the effective mechanical decay rate related to the linewidth of the OMIT window becomes a two-fold increase in the weak-coupling limit. When the dark-mode effect is broken, controllable double transparency windows appear and the second-order sideband, as well as the light delay or advance, is significantly enhanced. For an $N$-mechanical-mode optomechanical system, we find that in the presence of the dark-mode effect, the amplification multiple of the linewidth of the OMIT window is nearly proportional to the number of mechanical modes, and that the OMIT with a single window becomes the one with $N$ tunable windows by breaking the dark-mode effect. The study will be useful in optical information storage within a large frequency bandwidth and multi-channel optical communication based on optomechanical systems.
\end{abstract}

\maketitle

\section{Introduction}
Cavity optomechanical systems~\cite{Kippenberg2008Science,Meystre2013AP,Aspelmeyer2014RMP} are an important platform for manipulating mechanical properties through optical means and studying cavity-field statistics by mechanically changing the cavity boundary~\cite{Rabl2011PRL,Nunnenkamp2011,Liao2012PRA,Liao2013PRA,Liao2013,Wang2013PRL,Liu2013PRL,Vitali2007PRL,Genes2011PRA,Cirio2017PRL,Xu12015PRA,Wu2018PRApplied,Qin2018PRL,Lai2018PRA,Qin2019npj,Zippilli2018PRA,Barzanjeh2017NC}. Optomechanically induced transparency (OMIT)~\cite{Agarwal2010PRA,Weis2010Science,Safavi-Naeini2011Nature}, as a prominent application closely relevant to this platform, is the result of destructive interference between the anti-Stokes field and the probe field. The underlying  physical mechanism is analogous to electromagnetically induced transparency~\cite{Harris1990PRL,Harris1997PT}.
The performance of an OMIT process is mainly described by the optical transmission and delay, which will directly determine the information transfer efficiency and storage time, respectively. Recently, various schemes have been proposed to tune the transmission rate and group delay of the signal light, such as OMIT in hybridized optomechanical systems~\cite{Wang2014PRA,Ma2014PRA,Huang2014JPB,Hou2015PRA,Dong2015NC,Akram2015PRA,Jiang2016PRA,Shen2016OL,Zhang2018PRA,Lu2019PRA}, OMIT with higher-order sidebands~\cite{Xiong2012PRA,Kong2017PRA,Jiao2018PRA}, OMIT in quadratically nanomechanical systems~\cite{Huang2011PRA,Karuza2013PRA}, nonreciprocal OMIT~\cite{Shen2016NP}, OMIT with Bogoliubov mechanical modes~\cite{Dong2014OP}, and OMIT at exceptional points~\cite{Jing2016PRp}.

Though great advances have been made to improve OMIT, it remains a long-standing challenge to significantly broaden the linewidth of the OMIT window determining the frequency bandwidth of the information transmission and realize the ultra-long optical delay~\cite{Agarwal2010PRA,Weis2010Science,Safavi-Naeini2011Nature}. The physical origin behind this obstacle is that the effective mechanical decay rate relating to the linewidth of the OMIT window is governed by the pump laser power (i.e., the intracavity photon number)~\cite{Agarwal2010PRA,Weis2010Science,Safavi-Naeini2011Nature}, while a high pump power will cause the bistability of the system~\cite{Weis2010Science,Safavi-Naeini2011Nature,Xiong2012PRA,Wang2014PRA,Lu2019PRA}.

In parallel, the optomechanical dark mode~\cite{Dong2012Science,Wang2012PRL,Tian2012PRL}, proposed in a system of two optical cavities coupled to a common mechanical resonator, has attracted much attention in recent years. It can be employed for the realization of high efficient quantum state conversion~\cite{Wang2012PRL,Tian2012PRL} and the exploitation of circumventing the decay~\cite{Wang2012PRL,Tian2012PRL,Tian2013PRL,Wang2013PRL}. Meanwhile, two mechanical modes or magnon modes coupled to a common cavity field can also be hybridized into a bright mode and a dark mode decoupled from the system~\cite{Genes2008NJP,Shkarin2014PRL,Zhang2015NC,Kuzyk2017PRA,Sommer2019PRL,Ockeloen-Korppi2019PRA}. This physical mechanism has been broadly applied to gradient memory or information storage~\cite{Zhang2015NC}, and energy transfer~\cite{Shkarin2014PRL}. In particular, the effective mechanical decay rate of the bright mode is nearly twice that of the individual mechanical mode~\cite{Genes2008NJP,Shkarin2014PRL,Zhang2015NC,Kuzyk2017PRA,Sommer2019PRL,Ockeloen-Korppi2019PRA}. Thus, based on the characteristic that the narrow linewidth of the OMIT window results from the effective mechanical decay rate of the mechanical mode~\cite{Agarwal2010PRA,Weis2010Science,Safavi-Naeini2011Nature}, it is naturally to ask the question whether we can significantly widen the OMIT window or steer the switch from the OMIT with a single transparent window to the case of multiple transmission windows by controlling the dark-mode effect in a multi-mechanical-mode optomechanical system.

\begin{figure*}[tbp]
\center
\includegraphics[width=0.93 \textwidth]{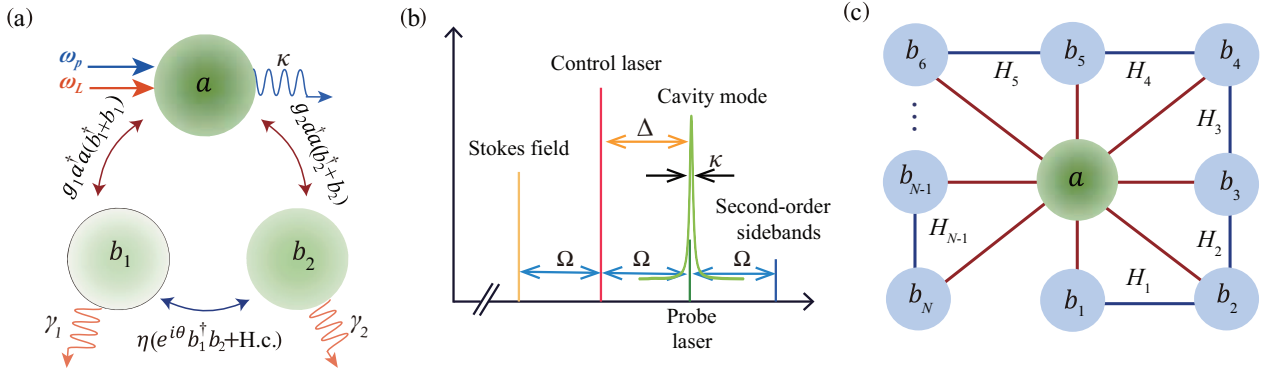}
\caption{(a) Schematic diagram of the three-mode loop-coupled optomechanical system formed by a cavity-field mode $a$ (with a decay rate $\kappa$) optomechanically coupled to two mechanical modes $b_{1}$ and $b_{2}$ (with decay rates $\gamma_{1}$ and $\gamma_{2}$). Two mechanical modes are coupled to each other via a phase-dependent phonon-phonon interaction with the coupling strength $\eta$ and the modulation phase $\theta$. The system is driven by a strong pump field of frequency $\omega_{L}$ and a weak probe field of frequency $\omega_{p}$.  (b) Frequency spectrum for the system shown in (a). The first-order sidebands with frequencies $\omega_{L}\pm\Omega$ are referred to as the Stokes and anti-Stokes fields, respectively. The sidebands with frequencies $\omega_{L}\pm 2\Omega$ are the second-order upper or lower sidebands. (c) We extend the scheme in (a) to a net-coupled optomechanical system: a cavity mode is simultaneously coupled to $N\geq 3$ ($N$ is an integer) mechanical modes via optomechanical couplings, and the nearest-neighboring mechanical modes are coupled to each other through the phase-dependent phonon-exchange couplings $H_{j}$ for $j=1$-$(N-1)$.}
\label{Figmodel}
\end{figure*}

In this paper, we study the sensitive impacts of the dark-mode effect on the optical properties, including the optical transmission, group delay, and higher-order sidebands. We consider a three-mode loop-coupled optomechanical system composed of a cavity-field mode optomechanically coupled to two mechanical modes, which are coupled to each other by a phase-dependent phonon-phonon interaction. Combining this phase-dependent phonon-phonon coupling with the optomechanical interactions, the quantum interference effect leads to the breaking of the dark mode. We find that in the presence of the dark mode, OMIT and second-order sidebands can be significantly enhanced compared to those in a standard optomechanical system, and the amplification multiple of the linewidth of the OMIT window is nearly proportional to the number of mechanical modes. This provides a route to achieve the optical transmission within a large frequency bandwidth. When breaking the dark-mode effect, the OMIT with a single window becomes one with multiple windows, and the transmission windows can be controlled and switched periodically by tuning the modulation phase. These results provide the possibility to enhance or steer the performance of OMIT, and realize the switch from single-channel to multichannel optical communication. In particular, the significantly enhanced second-order sideband and light delay or advance, which are potentially useful for the precise sensing~\cite{Nunnenkamp2013PRL,Xiong2017OL,Kong2017PRA,Li2013PR,Liu2017SR,Liu2018OL} and optical storage or quantum communication~\cite{Agarwal2010PRA,Weis2010Science,Safavi-Naeini2011Nature}, respectively, can be achieved by controlling the dark-mode effect.

The rest of this paper is organized as follows. In Sec.~\ref{sec2}, we describe the model and present the OMIT solution. In Sec.~\ref{sec3}, we analyze OMIT, second-order sideband, and group delay by using the dark-mode effect. In Sec.~\ref{sec4}, we extend our studies to the case of the $N$-mechanical-mode optomechanical system. Finally, we provide a brief conclusion in Sec.~\ref{sec5}. For keeping the completeness of this work, we present four Appendixes, in which we show some algebra equations and sideband parameters (Appendix A), the detailed calculation of the effective mechanical decay rate (Appendix B), and the detailed derivation of Eq.~(\ref{ck}) (Appendix C).

\section{Model and solutions \label{sec2}}

We consider a three-mode loop-coupled optomechanical system, which consists of an optical cavity mode coupled to two mechanical modes via radiation-pressure force, as shown in Fig.~\ref{Figmodel}(a). The two mechanical modes are coupled to each other via a phase-dependent phonon-phonon interaction with coupling strength $\eta$ and modulation phase $\theta$. Note that this model can be implemented in either photonic crystal optomechanical setups~\cite{Fang2017NPB} or circuit electromechanical systems ~\cite{Massel2011Nature,Massel2012Nc}. In the photonic crystal optomechanical systems, the phase-dependent phonon-exchange interaction has been suggested by using two assistant cavity fields~\cite{Fang2017NPB}. The phase-dependent phonon-exchange coupling in the circuit electromechanical setups can be indirectly induced by coupling the two mechanical modes to a charge qubit~\cite{Lai2020}. A pump laser with frequency $\omega_{L}$ and field amplitude $\varepsilon_{L}=\sqrt{2\kappa P_{L}/\hbar\omega_{L}}$, and a probe laser with frequency $\omega_{p}$ and amplitude $\varepsilon_{p}=\sqrt{2\kappa P_{p}/\hbar\omega_{p}}$ are applied to this system. Here, $P_{L}$ and $P_{p}$ are the control and probe light powers, respectively, and $\kappa$ is the cavity-field decay rate. In a rotating frame defined by the unitary transformation operator $\exp(-i\omega_{L}a^{\dagger}at)$, the system Hamiltonian reads (with $\hbar =1$)
\begin{equation}
\label{H}
H=H_{0}+H_{\text{int}}+H_{\text{dr}},
\end{equation}
with
\begin{subequations}
\label{H}
\begin{align}
H_{0}=&\Delta _{c}a^{\dagger }a+\sum_{l=1}^{2}\omega _{l}b_{l}^{\dagger}b_{l}, \\
H_{\text{int}}=&\sum_{l=1}^{2}g_{l}a^{\dagger }a(b_{l}+b_{l}^{\dagger}) +\eta(e^{i\theta}b_{1}^{\dagger}b_{2}+e^{-i\theta}b_{2}^{\dagger}b_{1}), \\
H_{\text{dr}}=&i(\varepsilon_{L}a^{\dagger }+\varepsilon_{p}a^{\dagger}e^{-i\Omega t}-\mathrm{H.c.}),
\end{align}
\end{subequations}
where $a$ ($a^{\dagger}$) and $b_{l}$ ($b^{\dagger}_{l}$) are, respectively, the annihilation (creation) operators of the cavity mode and the $l$-th mechanical mode, with the corresponding resonance frequencies $\omega_{c}$ and $\omega_{l}$. The optomechanical and the phonon-phonon interaction strengths are denoted by $g_{l}$ and $\eta$, respectively.
The cavity-pump and probe-pump detunings are given by
\begin{eqnarray}
\Delta_{c}=\omega_{c}-\omega_{L},\hspace{0.3 cm}\Omega=\omega_{p}-\omega_{L}.
\end{eqnarray}
When the control and probe lasers are simultaneously injected into the cavity, we consider the high-order sidebands with frequencies $\omega_{L}\pm n\Omega$ ($n$ represents the order of the sidebands) induced by the nonlinear optomechanical interaction. The frequency spectrum of the optomechanical system is shown in Fig.~\ref{Figmodel}(b), in which the first-order sidebands with frequencies $\omega_{L}\pm\Omega$ correspond to the anti-Stokes and Stokes fields and the sidebands with frequencies $\omega_{L}\pm 2\Omega$ denote the second-order upper or lower sideband, respectively. The Langevin equations for the operators of the optical and mechanical modes can be written as
\begin{subequations}
\label{Langevineqorig0}
\begin{align}
\dot{a}=&-(\kappa +i\Delta_{c})a-i\sum_{l=1,2}g_{l}a(b_{l}+b_{l}^{\dagger})\notag \\
&+\varepsilon_{L}+\varepsilon_{p}e^{-i\Omega t}+\sqrt{2\kappa}a_{\text{in}}, \\
\dot{b}_{1}=&-(\gamma _{1}+i\omega _{1})b_{1}-ig_{1}a^{\dagger }a-i\eta e^{i\theta }b_{2}+\sqrt{2\gamma_{1}}b_{1,\text{in}}, \\
\dot{b}_{2}=&-(\gamma _{2}+i\omega _{2})b_{2}-ig_{2}a^{\dagger }a-i\eta e^{-i\theta }b_{1}+\sqrt{2\gamma_{2}}b_{2,\text{in}},
\end{align}
\end{subequations}
where $\kappa$ and $\gamma_{l}$ are the decay rates of the cavity mode and the $l$-th mechanical mode, respectively. The operators $a_{\text{in}}$ and $b_{l,\text{in}}$ are, respectively, the noise operators of the cavity-field mode and the $l$th mechanical mode. We express the variables in Eq.~(\ref{Langevineqorig0}) as the sum of their steady-state values and quantum fluctuations, namely $a=\alpha+\delta a$ and $b_{l=1,2}=\beta_{l}+\delta b_{l}$.
As the control field is assumed to be sufficiently strong, all the operators can be identified with their expectation values~\cite{Weis2010Science,Xiong2012PRA}. Then, the equations for the expectation values of these fluctuations can be obtained as~\cite{Weis2010Science,Xiong2012PRA}
\begin{subequations}
\label{Langevinehigh}
\begin{align}
\delta \dot{a}=&-(\kappa+i\Delta) \delta a-\sum_{l=1,2}ig_{l}\alpha(\delta b_{l}+\delta b_{l}^{\dagger})\notag \\
&-\sum_{l=1,2}ig_{l}\delta a(\delta b_{l}+\delta b_{l}^{\dagger})+\varepsilon_{p}e^{-i\Omega t}, \\
\delta \dot{b}_{1}=&-(\gamma_{1}+i\omega _{1})\delta b_{1}-ig_{1}(\alpha ^{\ast}\delta a+\alpha \delta a^{\dagger})\notag \\
&-i\eta e^{i\theta }\delta b_{2}-ig_{1}\delta a^{\dagger}\delta a, \\
\delta \dot{b}_{2}=&-(\gamma _{2}+i\omega_{2})\delta b_{2}-ig_{2}(\alpha^{\ast }\delta a+\alpha\delta a^{\dagger})\notag \\
&-i\eta e^{-i\theta}\delta b_{1}-ig_{2}\delta a^{\dagger }\delta a,
\end{align}
\end{subequations}
where the quantum noise terms ($a_{\text{in}}$ and $b_{l=1,2,\text{in}}$) can be ignored due to their zero mean values. The parameter $\Delta =\Delta _{c}+\sum_{l=1,2}g_{l}(\beta_{l}+\beta_{l}^{\ast})$ denotes the effective detuning of the cavity field shifted by the optomechanical interactions, and the steady-state mean values are given by $\alpha=\varepsilon_{L}/(\kappa +i\Delta)$,   $\beta_{1}=-i(g_{1}\vert \alpha \vert ^{2}+\eta e^{i\theta }\beta _{2}) /(\gamma_{1}+i\omega_{1})$, and  $\beta_{2}=-i(g_{2}\vert \alpha
\vert ^{2}+\eta e^{-i\theta }\beta _{1}) /(\gamma _{2}+i\omega_{2})$. In Eq.~(\ref{Langevinehigh}), the nonlinear terms, such as $\delta a^{\dagger }\delta a$ and $\delta a(\delta b_{l}+\delta b_{l}^{\dagger})$, are kept to generate the required second-order sidebands, and the higher-order sidebands are neglected.
The response of the probe field for the first- and second-order sidebands to the system can be exhibited by using the following ansatz~\cite{Weis2010Science,Xiong2012PRA}
\begin{subequations}
\label{highorder}
\begin{align}
\delta a=&A_{1}^{-}e^{-i\Omega t}+A_{1}^{+}e^{i\Omega t}+A_{2}^{-}e^{-2i\Omega t}+A_{2}^{+}e^{2i\Omega t},  \\
\delta b_{1}=&B_{1,1}^{-}e^{-i\Omega t}+B_{1,1}^{+}e^{i\Omega t}+B_{1,2}^{-}e^{-2i\Omega t}+B_{1,2}^{+}e^{2i\Omega t}, \\
\delta b_{2}=&B_{2,1}^{-}e^{-i\Omega t}+B_{2,1}^{+}e^{i\Omega t}+B_{2,2}^{-}e^{-2i\Omega t}+B_{2,2}^{+}e^{2i\Omega t}.
\end{align}
\end{subequations}
By substituting Eq.~(\ref{highorder}) into Eq.~(\ref{Langevinehigh}), we obtain twelve algebraic equations which are divided into two groups, as shown in Appendix~\ref{appendixa}. After solving Eqs.~(\ref{firsteq}) and ~(\ref{secondeq}), we obtain the coefficients of the first- and second-order upper sidebands, which can describe the linear and nonlinear responses of the system.
\begin{widetext}
The coefficient of the first-order upper sideband is
\begin{equation}
\label{firstorder}
A_{1}^{-}=\frac{T_{2}^{(1)}[-\kappa +i(\Delta +\Omega )]-2i\left\vert \alpha\right\vert^{2}(g_{2}^{2}T_{3,1}^{(1)}+g_{1}^{2}T_{3,2}^{(1)})-4iT_{1}^{(1)}g_{1}g_{2}\eta \left\vert \alpha \right\vert ^{2}\cos \theta }{-T_{2}^{(1)}[\Delta^{2}+\left( \kappa -i\Omega \right) ^{2}]+4\left\vert \alpha \right\vert^{2}\Delta(g_{2}^{2}T_{3,1}^{(1)}+g_{1}^{2}T_{3,2}^{(1)})+8T_{1}^{(1)}g_{1}g_{2}\eta\left\vert \alpha \right\vert ^{2}\Delta \cos \theta }\varepsilon_{p},
\end{equation}
and that of the second-order upper sideband is
\begin{equation}
\label{secondorder}
A_{2}^{-}=\frac{\chi _{1}(\Omega )(A_{1}^{+})^{\ast
}+i[g_{1}(B_{1,1}^{-}+(B_{1,1}^{+})^{\ast
})+g_{2}(B_{2,1}^{-}+(B_{2,1}^{+})^{\ast })]A_{1}^{-}}{\chi _{2}(\Omega
)-[\kappa +i(\Delta -2\Omega )]},
\end{equation}
with
\begin{eqnarray}
\label{paremeters1}
(A_{1}^{+})^{\ast } &=&\frac{-2\left( \alpha ^{\ast }\right) ^{2}\left[\kappa -i\left( \Delta +\Omega \right) \right] \left[ 2g_{1}g_{2}\eta T_{1}^{(1)}\cos \theta +\left(g_{1}^{2}T_{3,2}^{(1)}+g_{2}^{2}T_{3,1}^{(1)}\right) \right] }{\left(i\kappa +\Delta +\Omega \right) \left\{ \left[ 4\left\vert \alpha\right\vert ^{2}\Delta \left(g_{2}^{2}T_{3,1}^{(1)}+g_{1}^{2}T_{3,2}^{(1)}\right) -T_{2}^{(1)}\left[\Delta ^{2}+\left( \kappa -i\Omega \right) ^{2}\right] \right]+8g_{1}g_{2}\eta \left\vert \alpha \right\vert ^{2}T_{1}^{(1)}\Delta \cos\theta \right\} }\varepsilon_{p}, \\
\chi _{1}(\Omega ) &=&\frac{2i\alpha \left[ g_{1}\alpha \left( \left(B_{1,1}^{+}\right) ^{\ast }+B_{1,1}^{-}\right) +g_{2}\alpha \left( \left(B_{2,1}^{+}\right) ^{\ast }+B_{2,1}^{-}\right) -A_{1}^{-}\left( i\kappa
+\Delta +2\Omega \right) \right] \left[ 2g_{1}g_{2}\eta T_{1}^{(2)}\cos \theta +\left( g_{1}^{2}T_{3,2}^{(2)}+g_{2}^{2}T_{3,1}^{(2)}\right) \right]}{\left[ \left( i\kappa +\Delta +2\Omega \right) T_{2}^{(2)}-2\left\vert\alpha \right\vert ^{2}\left(g_{1}^{2}T_{3,2}^{(2)}+g_{2}^{2}T_{3,1}^{(2)}\right) \right]-4g_{1}g_{2}\eta \left\vert \alpha \right\vert ^{2}T_{1}^{(2)}\cos \theta },
\\
\chi _{2}(\Omega ) &=&\left[ [\kappa -i(\Delta +2\Omega)]^{-1}-iT_{2}^{(2)}[2\left\vert \alpha \right\vert^{2}(g_{1}^{2}T_{3,2}^{(2)}+g_{2}^{2}T_{3,1}^{(2)}+2g_{1}g_{2}\eta
T_{1}^{(2)}\cos \theta )]^{-1}\right] ^{-1}.
\end{eqnarray}
\end{widetext}
Other parameters used are displayed in Appendix~\ref{appendixa}. In the following, by using the input-output relation~\cite{Gardiner1985PRA}, we obtain the transmission rate of the probe field and the efficiency of the second-order sideband~\cite{Weis2010Science,Safavi-Naeini2011Nature,Xiong2012PRA,Jiao2018PRA}:
\begin{equation}
\vert t_{p}\vert ^{2}=\left\vert 1-\frac{\kappa }{\varepsilon_{p}}A_{1}^{-}\right\vert ^{2},\hspace{0.3 cm}\Lambda_{p}=\left\vert -\frac{\kappa }{\varepsilon_{p}}A_{2}^{-}\right\vert.
\end{equation}
The associated transmission group delay caused by the rapid phase dispersion is given by~\cite{Safavi-Naeini2011Nature}
\begin{align}
\label{groupdelay}
\tau_{1}=\frac{d\arg(t_{p})}{d\Omega}|_{\Omega=\omega_{l}},
\end{align}
where $\arg(x)$ takes the argument of the complex number $x$.
\section{OMIT by controlling the dark mode\label{sec3}}

\subsection{Dark-mode effect and its breaking \label{darkmode}}

In this section, we analyze the dark-mode effect and its breaking in the three-mode optomechanical system. In the strong-driving regime, the linearized optomechanical Hamiltonian can be inferred from the equations of motion for the quantum fluctuations in Eq.~(\ref{Langevinehigh}). Under the rotating-wave approximation (RWA), the linearized optomechanical Hamiltonian takes the form
\begin{eqnarray}
H_{\text{RWA}}&=&\Delta \delta a^{\dagger}\delta a+\omega_{1}\delta b_{1}^{\dagger}\delta b_{1}+\omega_{2}\delta b_{2}^{\dagger}\delta b_{2}+\eta(e^{i\theta}\delta b_{1}^{\dagger}\delta b_{2}\nonumber \\
&&+e^{-i\theta}\delta b_{2}^{\dagger}\delta b_{1})+G_{1}(\delta a\delta b_{1}^{\dagger}+\delta b_{1}\delta a^{\dagger})\nonumber \\
&&+G_{2}(\delta a\delta b_{2}^{\dagger}+\delta b_{2}\delta a^{\dagger}),\label{RWAH}
\end{eqnarray}
where $\Delta$ is the effective detuning of the cavity field shifted by the optomechanical interactions, and $G_{l}=g_{l}|\alpha|$ is the linearized optomechanical coupling strength.

To investigate the dark-mode effect in this system, we first consider the case where the phase-modulated phonon-exchange interaction is absent, i.e., $\eta=0$. In this case, due to an optically mediated coupling, the two mechanical modes are hybridized into a bright mode and a dark mode, which are expressed as
\begin{eqnarray}
B_{+}&=&\frac{1}{\sqrt{G^{2}_{1}+G^{2}_{2}}}(G_{1}\delta b_{1}+G_{2}\delta b_{2}),\\
B_{-}&=&\frac{1}{\sqrt{G^{2}_{1}+G^{2}_{2}}}(G_{2}\delta b_{1}-G_{1}\delta b_{2}).
\end{eqnarray}
Then, the Hamiltonian in Eq.~(\ref{RWAH}) can be rewritten with the two hybridization modes $B_{\pm}$ as
\begin{eqnarray}
H_{\text{Hyb}} &=&\Delta\delta a^{\dagger }\delta a+\omega_{+}B_{+}^{\dagger}B_{+}+\omega_{-}B_{-}^{\dagger }B_{-}+\zeta(B_{+}^{\dagger}B_{-}+B_{-}^{\dagger}B_{+})\nonumber \\
&&+G_{+}(\delta aB_{+}^{\dagger}+B_{+}\delta a^{\dagger }),\label{Parity}
\end{eqnarray}
where we introduce these parameters
\begin{eqnarray}
\omega_{+} &=&\frac{G^{2}_{1}\omega_{1}+G^{2}_{2}\omega_{2}}{G^{2}_{1}+G^{2}_{2}},\hspace{0.5 cm}
\omega_{-} =\frac{G^{2}_{2}\omega_{1}+G^{2}_{1}\omega_{2}}{G^{2}_{1}+G^{2}_{2}}, \\
\zeta &=&\frac{G_{1}G_{2}(\omega_{1}-\omega_{2})}{G^{2}_{1}+G^{2}_{2}}, \hspace{0.5 cm}
G_{+}=\sqrt{G^{2}_{1}+G^{2}_{2}}.\label{ParityG}
\end{eqnarray}
It can be seen from Eq.~(\ref{ParityG}) that when $\omega_{1}=\omega_{2}$, we have $\zeta=0$, and thus the mode $B_{-}$ is decoupled from both the cavity mode $a$ and the mode $B_{+}$, which means that the dark mode $B_{-}$ appears in this system.

To break the dark-mode effect, a phase-dependent phonon-exchange interaction between the two mechanical modes is considered. Then, by introducing two new bosonic modes $\tilde{B}_{+}$ and $\tilde{B}_{-}$ defined as
\begin{eqnarray}
\tilde{B}_{+}&=&f\delta b_{1}-e^{i\theta}h\delta b_{2},\hspace{0.5 cm}
\tilde{B}_{-}=e^{-i\theta}h\delta b_{1}+f\delta b_{2},
\end{eqnarray}
Hamiltonian~(\ref{RWAH}) becomes
\begin{eqnarray}
H_{\text{D}}&=&\Delta \delta a^{\dagger }\delta a+\tilde{\omega}_{+}\tilde{B}_{+}^{\dagger}\tilde{B}_{+}+\tilde{\omega}_{-}\tilde{B}_{-}^{\dagger}\tilde{B}_{-} +(\tilde{G}_{+}^{\ast}\delta a\tilde{B}_{+}^{\dagger}+\tilde{G}_{+}\tilde{B}_{+}\delta a^{\dagger})\nonumber \\
&&+(\tilde{G}_{-}^{\ast}\delta a\tilde{B}_{-}^{\dagger}+\tilde{G}_{-}\tilde{B}_{-}\delta a^{\dagger}),\label{DigH}
\end{eqnarray}
where the coupling strengths and resonance frequencies are given by
\begin{eqnarray}
\label{DigG}
\tilde{G}_{+}&=&fG_{1}-e^{-i\theta }hG_{2},\hspace{0.5 cm}
\tilde{G}_{-}=e^{i\theta }hG_{1}+fG_{2},\nonumber\\
\tilde{\omega}_{\pm}& =&\frac{1}{2}\left(\omega _{1}+\omega _{2}\pm \sqrt{(\omega_{1}-\omega_{2})^{2}+4\eta ^{2}}\right),
\end{eqnarray}
with $f={\vert\tilde{\omega}_{-}-\omega_{1}\vert}/\sqrt{(\tilde{\omega}_{-}-\omega_{1})^{2}+\eta^{2}}$ and $h=\eta f/(\tilde{\omega}_{-}-\omega_{1})$. When the two mechanical modes have the same frequencies, $\omega_{1}=\omega_{2}=\omega_{m}$, and coupling strengths, $G_{1}=G_{2}=G$, the coupling strengths in Eq.~(\ref{DigG}) can be simplified as
\begin{eqnarray}
\label{DigGequwG12}
\tilde{G}_{+}&=&\sqrt{2}G(1+e^{-i\theta })/2,\hspace{0.5 cm}
\tilde{G}_{-}=\sqrt{2}G(1-e^{i\theta })/2.
\end{eqnarray}
It follows from Eq.~(\ref{DigGequwG12}) that, when $\theta=n\pi$ for an integer $n$, the cavity mode $a$ is decoupled from one of the two hybridized mechanical modes $\tilde{B}_{-}$ (for an even number $n$) and $\tilde{B}_{+}$ (for an odd number $n$). These features mean that the dark-mode effect can be broken by tuning the modulation phase $\theta\neq n\pi$.

\begin{figure*}[tbp]
\centering
\includegraphics[width=0.9 \textwidth]{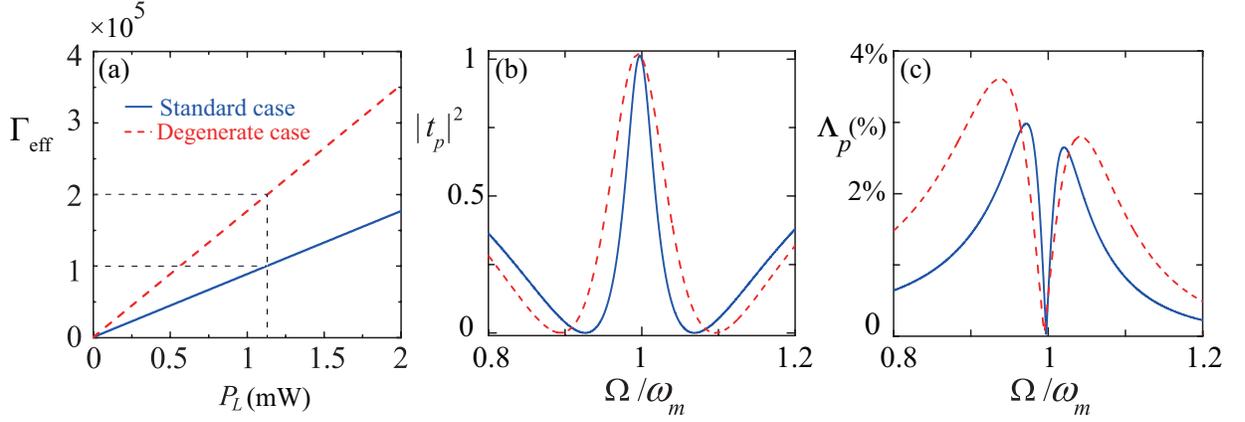}
\caption{(Color online) In the presence of the dark-mode effect ($\eta=0$), (a) the linewidth $\Gamma_{\text{eff}}$ of the OMIT window as a function of the pump power $P$ for a standard optomechanical system ($g_{2}=0$, blue solid lines) and a two-degenerate-mechanical-mode optomechanical system ($g_{1}=g_{2}$, red dashed lines). (b) The transmission rate $|t_{p}|^{2}$ of the probe light, and (c) the efficiency $\Lambda_{p}$($\%$) of the second-order sideband as a function of the probe-pump detuning $\Omega$ when $P_{L}=1.5$ mW for a standard optomechanical system (blue solid lines) and a two-degenerate-mechanical-mode optomechanical system (red dashed lines). Here we take the normalized cavity-pump detuning $\Delta/\omega_{m}=1$.}\label{singledouble}
\end{figure*}

\subsection{OMIT and second-order sidebands in the presence of the dark mode\label{dark-OMIT}}

In the above sections, we have derived the transmission rate of the probe field and the efficiency of the second-order sideband, and have analyzed how to control the dark-mode effect in the two-mechanical-mode optomechanical system. Now, we study OMIT and its second-order sidebands in the presence of the dark mode ($\eta=0$). To make the results feasible in experiments, we use the parameters realized in recent experiments~\cite{Hill2012NC,Weaver2017NC}, i.e., $\lambda=1064$ nm, $L=25$ mm, $\kappa=2\pi\times215$ kHz, $\omega_{l=1,2}=\omega_{m}=2\pi\times947$ kHz, $m_{l=1,2}=145$ ng, $Q_{l}=6700$ ($\gamma_{l}=\omega_{l}/Q_{l}$), and $\varepsilon_{p}=0.05\varepsilon_{L}$.

A single OMIT is due to the destructive interference between the probe field and the anti-Stokes scattering stimulated by the red-sideband driving ~\cite{Agarwal2010PRA,Weis2010Science,Safavi-Naeini2011Nature}. Correspondingly, in the case of the two mechanical modes, there exist two routes of destructive interferences which lead to two transparency windows in the OMIT spectrum. However, we find some counterintuitive phenomena in the presence of the dark mode ($\eta=0$) shown in Fig.~\ref{singledouble}. There exists only one transparency window and one second-order sideband window in the two-degenerate-mechanical-mode optomechanical system [see the red dashed lines in Figs.~\ref{singledouble}(b) and ~\ref{singledouble}(c)]. Physically, this is due to the fact that the bright mode plays an effective role in the destructive interference and the dark mode is decoupled from the system. Moreover, compared to a standard optomechanical system consisting of a cavity mode and a single mechanical mode, it exhibits not only a multifold amplified OMIT window but also a significantly enhanced second-order sideband in the two-degenerate-mechanical-mode optomechanical system. In a standard optomechanical system, the width of the OMIT window is related to the effective mechanical decay rate given by~\cite{Agarwal2010PRA,Weis2010Science,Safavi-Naeini2011Nature}
\begin{eqnarray}
\label{effectdecay1}
\Gamma_{\text{eff}}=\gamma_{m}+\gamma_{\text{opt}},
\end{eqnarray}
where $\gamma_{\text{opt}}\approx G^{2}/\kappa$ stands for the optomechanically induced mechanical decay rate for a single mechanical mode (with the mechanical decay rate $\gamma_{m}$), and $G=g|\alpha|$ ($|\alpha|^{2}$ is the intracavity photon number) denotes the linearized optomechanical coupling strength.

For the two-degenerate-mechanical-mode optomechanical system, we can also obtain the effective mechanical damping rate of the bright mode by adiabatically eliminating the cavity field in the large-cavity-decay regime (see Appendix~\ref{appendixb}), which is given by
\begin{eqnarray}
\label{effectdecay2}
\Gamma_{\text{eff}} &=&\gamma_{m}+2\gamma_{\text{opt}}.
\end{eqnarray}
Because of the parameter relation $\gamma_{\text{opt}}\gg\gamma_{m}$~\cite{Wilson-Rae2007PRL,Marquardt2007PRL,Genes2008PRA,Li2008PRB} in the weak-coupling regimes, we can see from Eq.~(\ref{effectdecay2}) that the linewidth of the OMIT window is approximately twice amplified in comparison with the case of the standard optomechanical system [see Eq.~(\ref{effectdecay1})]. Correspondingly, we plot the linewidth $\Gamma_{\text{eff}}$ of the OMIT window as a function of the pump power $P$, as shown in Fig.~\ref{singledouble}(a). On the one hand, the linewidth $\Gamma_{\text{eff}}$ of the OMIT window is significantly widened with the increase of the pump power $P$ in both the degenerate-mechanical-mode and prototype cases. On the other hand, the OMIT window in the degenerate case [see the red dashed line in Fig.~\ref{singledouble}(a)] is approximately twofold amplified in comparison with the prototype case [see the blue solid line in Fig.~\ref{singledouble}(a)] due to the existence of the dark-mode effect. This physical mechanism can be potentially used in the optical information storage within a large frequency bandwidth.

\begin{figure}[tbp]
\centering
\includegraphics[bb=0 0 331 335, width=0.47 \textwidth]{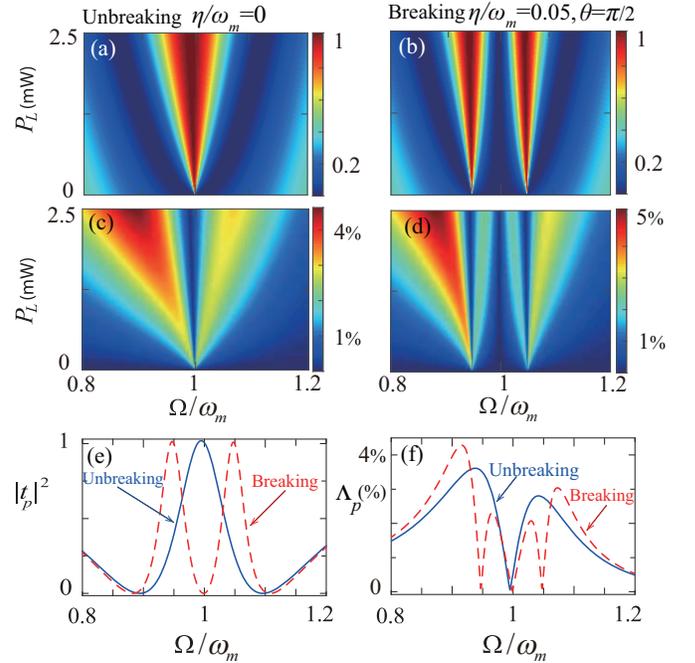}
\caption{(Color online) (a), (b) The transmission rate $|t_{p}|^{2}$ of the probe light and (c), (d) the efficiency $\Lambda_{p}$($\%$) of the second-order sideband as a function of the probe-pump detuning $\Omega$ and the pump light power $P_{L}$ in both the dark-mode-unbreaking ($\eta=0$) and -breaking ($\eta=0.05\omega_{m}$ and $\theta=\pi/2$) cases. (e) $|t_{p}|^{2}$ and (f) $\Lambda_{p}$($\%$) as functions of $\Omega$ when $P_{L}=1.5$ mW. Other parameters used are $g_{1}=g_{2}$ and $\Delta/\omega_{m}=1$.}\label{Pdetuning}
\end{figure}

\subsection{OMIT and the second-order sidebands by breaking the dark mode\label{break-OMIT}}

Since the above counterintuitive results are due to the dark-mode effect, it is natural to ask the question whether we can break the dark-mode effect to further explore the OMIT and the second-order sidebands. Thus, we compare the dark-mode-breaking case ($\eta=0.05\omega_{m}$ and $\theta=\pi/2$) with the dark-mode-unbreaking case ($\eta=0$), as shown in Fig.~\ref{Pdetuning}. In the dark-mode-unbreaking case, a single transparency window emerges around $\Omega=\omega_{m}$, when $\Delta=\omega_{l=1,2}=\omega_{m}$ [see Fig.~\ref{Pdetuning}(a)].

In contrast, when switching to the dark-mode-breaking case, a single OMIT window with linewidth $(\gamma_{m}+2\gamma_{\text{opt}})$ is divided into two symmetrical narrow OMIT windows around $\Omega\approx0.95\omega_{m}$ and $\Omega\approx1.05\omega_{m}$ with the linewidth $(\gamma_{m}+\gamma_{\text{opt}})$. Physically, the two degenerate mechanical modes are dressed into two well-separated dressed modes by breaking the dark-mode effect (see Sec.~\ref{darkmode}), and there exist the two routes of destructive interference which result in the two transparency windows even in the degenerate-mechanical-mode optomechanical system. Meanwhile, the second-order sideband shows that a single local minimum value is tuned to two symmetrical minimum values around $\Omega\approx0.95\omega_{m}$ and $\Omega\approx1.05\omega_{m}$. This is because the anti-Stokes field is resonantly enhanced, which promotes OMIT and leads to the suppression of the second-order sidebands.

\begin{figure}[tbp]
\centering
\includegraphics[bb=11 5 326 266, width=0.47 \textwidth]{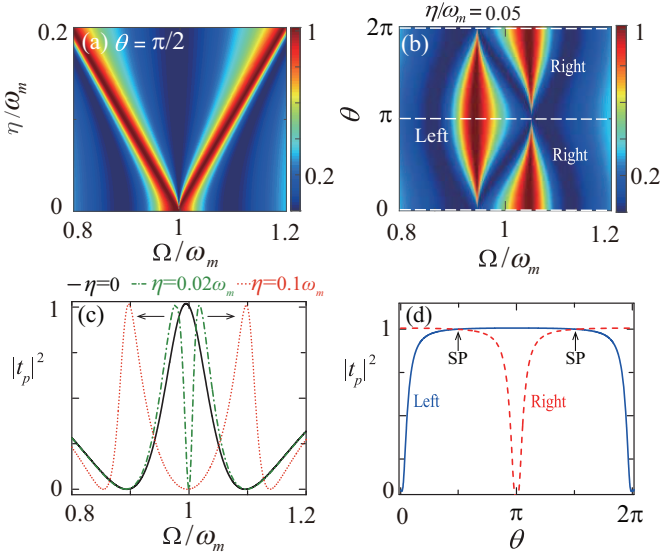}
\caption{(Color online) (a) The transmission rate $|t_{p}|^{2}$ of the probe light as a function of the phonon-phonon coupling $\eta$ and the probe-pump detuning $\Omega$ when $\theta=\pi/2$. (b) $|t_{p}|^{2}$ versus $\theta$ and $\Omega$ when $\eta/\omega_{m}=0.05$. The white dashed lines correspond to $\theta=0$, $\pi$ and $2\pi$. (c) $|t_{p}|^{2}$ as a function of $\Omega$ when $\eta$ takes different values: $\eta/\omega_{m}=0$ (black solid line), $0.02$ (green dashed-dotted line), and $0.1$ (red dotted line), under $\theta=\pi/2$. (d) $|t_{p}|^{2}$ versus $\theta$ at the two transparency windows $\Omega/\omega_{m}=0.95$ (left) and $\Omega/\omega_{m}=1.05$ (right), when $\eta/\omega_{m}=0.05$. Here, the switch points (SP) stand for the cross between the left and right transmission rates. Here we choose $P_{L}=1.5$ mW, $g_{1}=g_{2}$, and $\Delta/\omega_{m}=1$.}\label{etatheta}
\end{figure}

Now, we investigate the sensitive effects of the phonon-phonon coupling strength $\eta$ and the modulation phase $\theta$ on the transmission rates of the probe light and second-order sidebands. It is shown in Figs.~\ref{etatheta}(a) and ~\ref{etatheta}(c) that the switching between a single wide transparency window and two narrow transparency windows can be realized by tuning $\eta$ for $\theta=\pi/2$. The distance of the two OMIT windows is mainly determined by the coupling strength between the two phonon modes~\cite{Ma2014PRA}. Meanwhile, the transmission rate of the probe light depends on $\theta$, as shown in Figs.~\ref{etatheta}(b) and ~\ref{etatheta}(d). In the region $0<\theta<\pi$ ($\pi<\theta<2\pi$), the left (right) OMIT window always becomes much deeper and broader while the right (left) one becomes weaker with the modulation phase $\theta$, e.g., the right (left) transparency even becomes completely absorbed by decreasing the phase down near $\theta=\pi$ ($2\pi$).

\begin{figure}[tbp]
\centering
\includegraphics[bb=3 16 335 285, width=0.47 \textwidth]{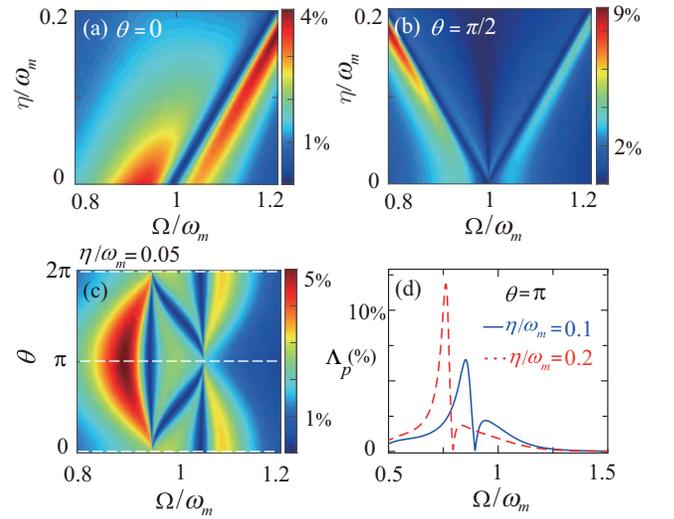}
\caption{(Color online) The efficiency $\Lambda_{p}$($\%$) of the second-order sideband as a function of the phonon-phonon coupling $\eta$ and the probe-pump detuning $\Omega$ when the modulation phase takes different
values (a) $\theta=0$ and (b) $\theta=\pi/2$. (c) $\Lambda_{p}$($\%$) versus $\theta$ and $\Omega$ when $\eta/\omega_{m}=0.05$; The white dashed lines correspond to $\theta=0$, $\pi$ and $2\pi$. (d) $\Lambda_{p}$($\%$) versus $\Omega$ when $\eta$ takes different values: $\eta/\omega_{m}=0.1$ and 0.2, for $\theta=\pi$. Here we consider $P_{L}=1.5$ mW, $g_{1}=g_{2}$, and $\Delta/\omega_{m}=1$.}\label{highorderetatheta}
\end{figure}

\begin{figure*}[tbp]
\centering
\includegraphics[width=0.93 \textwidth]{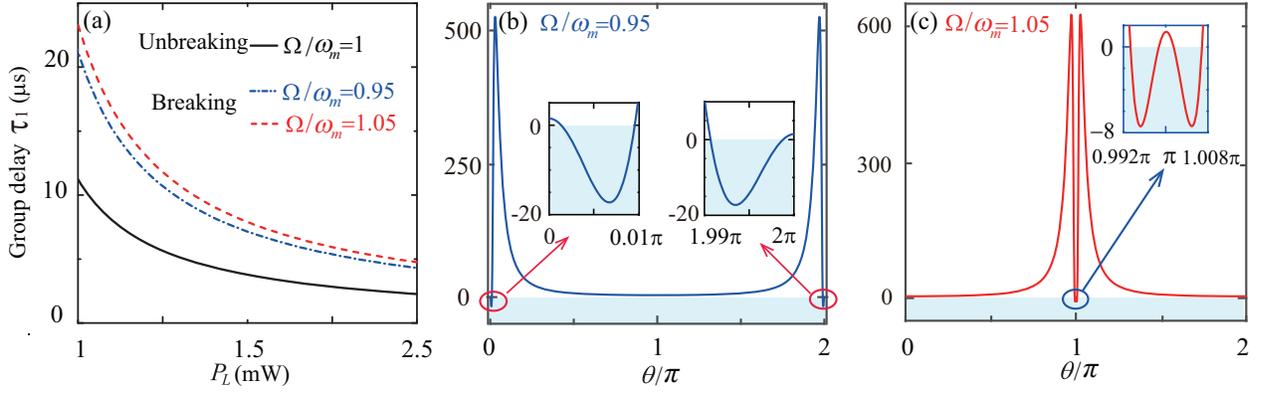}
\caption{(Color online) (a) Optical group delay $\tau_{1}$ as a function of the applied pump power $P_{L}$ for the dark-mode-unbreaking case ($\eta=0$) at its transmission window $\Omega/\omega_{m}=1$ (the black solid line) and the dark-mode-breaking case ($\eta=0.05\omega_{m}$ and $\theta=\pi/2$) at the left OMIT window $\Omega=0.95\omega_{m}$ (the blue dashed-dotted line) and the right OMIT window $\Omega=1.05\omega_{m}$ (the red dashed line), respectively. (b) At the left OMIT window $\Omega=0.95\omega_{m}$ and (c) the right OMIT window $\Omega=1.05\omega_{m}$, $\tau_{1}$ versus $\theta$ when $\eta=0.05\omega_{m}$ and $P_{L}=1.5$ mW. The insets in (b) and (c) are the zoomed-in plots of $\tau_{1}$ as a function of the modulation phase $\theta$. Here, we consider $g_{1}=g_{2}$ and $\Delta/\omega_{m}=1$.}\label{slowfast}
\end{figure*}

Furthermore, it is shown in Fig.~\ref{etatheta}(d) how the OMIT at the two transparency points $\Omega/\omega_{m}=0.95$ (left) and $\Omega/\omega_{m}=1.05$ (right) is modulated by the phase $\theta$. There are two switch points (SP) (i.e., the symmetrical transparency points $|t_{p}|_{\text{left}}^{2}=|t_{p}|_{\text{right}}^{2}$) when the modulation phase is given by $\theta=\pi/2$ and $3\pi/2$. It obviously shows that the OMIT performance of the left window (the blue solid line) is better than that of the right one (the red dashed line) between the two SP (i.e., $\pi/2<\theta<3\pi/2$), while the opposite situation that the OMIT performance of the right one (the red dashed line) is better than that of the left window (the blue solid line) appears in the rest region (i.e., $0<\theta<\pi/2$ and $3\pi/2<\theta<2\pi$). Moreover, one transparency window is lost in the presence of the dark mode ($\theta=n\pi$ for an integer $n$). Thus, the periodically controllable and switchable OMIT can be performed by tuning the modulation phase $\theta$. Physically, the combination of the $\theta$-dependent phonon-phonon interaction with the optomechanical couplings breaks the dark-mode effect and splits the OMIT spectrum with a single window.

Correspondingly, the dependence of the second-order sideband on the $\theta$-dependent phonon-phonon interaction is shown in Fig.~\ref{highorderetatheta}. We can see from Fig.~\ref{highorderetatheta}(a) that due to the existence of the dark-mode effect for $\theta=0$ (see Sec.~\ref{darkmode}), there always exists only one local minimum window even with the increase of the phonon-phonon coupling $\eta$.
When $\theta=\pi/2$ in Fig.~\ref{highorderetatheta}(b), the local minimum window is split into two windows in the second-order sideband by tuning $\eta$, in which the second-order sideband is significantly enhanced. For example, the maximum efficiency of the second-order sidebands is about $9\%$ [Fig.~\ref{highorderetatheta}(b)], which is much higher than that of the case $\theta=0$ [Fig.~\ref{highorderetatheta}(a)]. The $\theta$-dependent second-order sideband is exhibited in Fig.~\ref{highorderetatheta}(c) when $\eta/\omega_{m}=0.05$. We can see from Fig.~\ref{highorderetatheta}(c) that in the region $0<\theta<\pi$ ($\pi<\theta<2\pi$), the maximum efficiency becomes much larger (weaker). Especially, the highest efficiency emerges when the modulation phase takes $\theta=\pi$. This indicates that the second-order sidebands shown in Fig.~\ref{highorderetatheta}(d) can be significantly enhanced by increasing $\eta$ when $\theta=\pi$ [e.g., $\Lambda_{p}$ is about $12.5\%$, which is three times larger than that of $\theta=0$ shown in Fig.~\ref{highorderetatheta}(a)]. The significant enhancement of the second-order sidebands is important for its potential applications in weak-signal sensing~\cite{Nunnenkamp2013PRL,Xiong2017OL,Kong2017PRA,Li2013PR,Liu2017SR,Liu2018OL}, e.g., precise sensing of weak forces~\cite{Nunnenkamp2013PRL} and charges~\cite{Xiong2017OL,Kong2017PRA}.

\subsection{Controllable group delay\label{group delay}}

In general, the optical group delay can be created due to the fact that the dispersion curve varies drastically with the frequency within the OMIT window~\cite{Safavi-Naeini2011Nature}, which can be useful in optical information storage without absorption. In the dark-mode-unbreaking case ($\eta=0$), the slow-light effect emerges at its transmission window $\Delta=\omega_{m}$ [the black solid line in Fig.~\ref{slowfast}(a)]. When breaking the dark-mode effect (e.g., $\eta=0.05\omega_{m}$ and $\theta=\pi/2$), the group delay time significantly increases at both the left transmission window $\Delta=0.95\omega_{m}$ [the blue dash-dotted line in Fig.~\ref{slowfast}(a)] and the right transmission window $\Delta=1.05\omega_{m}$ [the red dashed line in Fig.~\ref{slowfast}(a)] in comparison with the dark-mode-unbreaking case. This implies that we can significantly develop the storage of the signal light by breaking the dark-mode effect.

In the dark-mode-breaking case, we also plot the optical group delay $\tau_{1}$ as a function of the modulation phase $\theta$ at the left transmission window $\Delta=0.95\omega_{m}$ [the blue solid line in Fig.~\ref{slowfast}(b)] and the right transmission window $\Delta=1.05\omega_{m}$ [the red solid line in Fig.~\ref{slowfast}(c)]. The slow-to-fast and fast-to-slow light effects can emerge at the two OMIT windows by tuning the modulation phase $\theta$ [see the insets in Figs.~\ref{slowfast}(b) and ~\ref{slowfast}(c)]. The delay time of the signal light can even reach 520 $\mathrm{\mu s}$ for the left window [see Fig.~\ref{slowfast}(b)] and 620 $\mathrm{\mu s}$ for the right window [see Fig.~\ref{slowfast}(c)], which is approximatively 100 times longer for the group delay in comparison with the dark-mode-unbreaking case. These results can lead to achieving ultra-slowing or ultra-advancing signals, which can be used in optical storage or quantum communication~\cite{Agarwal2010PRA,Weis2010Science,Safavi-Naeini2011Nature}.

\section{Tunable OMIT in an $N$-mechanical-mode optomechanical system\label{sec4}}

In this section, we extend our scheme to investigate the OMIT in an $N$-mechanical-mode optomechanical system, in which a cavity mode is optomechanically coupled to $N\geq 3$ (for an integer $N$) mechanical modes. And the neighboring mechanical modes are coupled to each other through the phase-dependent phonon-phonon interactions $H_{j}$ for $j=1$-$(N-1)$ with the coupling strength $\eta_{j}$ and modulation phase $\theta_{j}$ [see Fig.~\ref{Figmodel}(c)]. Thus, the Hamiltonian of the net-coupled optomechanical system can be written, in a frame rotating at the driving frequency, as
\begin{eqnarray}
H_{I}&=&\Delta _{c}a^{\dagger }a+\sum_{j=1}^{N}\omega _{j}b_{j}^{\dagger}b_{j}+\sum_{j=1}^{N}g_{j}a^{\dagger }a(b_{j}+b_{j}^{\dagger})\nonumber \\
&&+i(\varepsilon_{L}a^{\dagger }+\varepsilon_{p}a^{\dagger}e^{-i\Omega t}-\mathrm{H.c.})+\sum_{j=1}^{N-1}H_{j},
\end{eqnarray}
with
\begin{eqnarray}
H_{j}=\eta _{j}(e^{i\theta_{j}}b_{j}^{\dagger}b_{j+1}+e^{-i\theta_{j}}b_{j+1}^{\dagger}b_{j}),
\end{eqnarray}
where $\Delta_{c}=\omega_{c}-\omega_{L}$ is the detuning of the cavity frequency $\omega_{c}$ with respect to the driving frequency $\omega_{L}$.  We consider the strong driving case of the cavity and then perform the linearization procedure in this system. Meanwhile, the couplings between the cavity field and these mechanical modes are much smaller than the mechanical frequency and hence the RWA is justified. In this case, the linearized optomechanical Hamiltonian under
the RWA is given by
\begin{eqnarray}
H_{\text{RWA}}&=&\Delta \delta a^{\dagger}\delta a+\sum_{j=1}^{N}\omega _{j}\delta b_{j}^{\dagger}\delta b_{j}+\sum_{j=1}^{N}G_{j}(\delta a^{\dagger}\delta b_{j}+\delta b_{j}^{\dagger}\delta a)\nonumber\\
&&+\sum_{j=1}^{N-1}\eta_{j}(e^{i\theta_{j}}\delta b_{j}^{\dagger}\delta b_{j+1}+e^{-i\theta_{j}}\delta b_{j+1}^{\dagger}\delta b_{j}), \label{eq1iniH1Nm}
\end{eqnarray}
where $\Delta =\Delta_{c}+\sum_{j=1}^{N}g_{j}(\beta_{j}+\beta_{j}^{\ast})$ is the normalized driving detuning after the linearization, and $G_{j}=g_{j}|\alpha|$ is the linearized optomechanical coupling strength between the $j$th mechanical mode and the cavity-field mode. For convenience, we consider the case where all the mechanical modes have the same resonance frequencies ($\omega_{j}=\omega_{m}$), optomechanical coupling strengths ($G_{j}=G$), and phonon-exchange coupling strengths ($\eta_{j}=\eta$).

\begin{figure*}[tbp]
\centering
\includegraphics[width=0.93\textwidth]{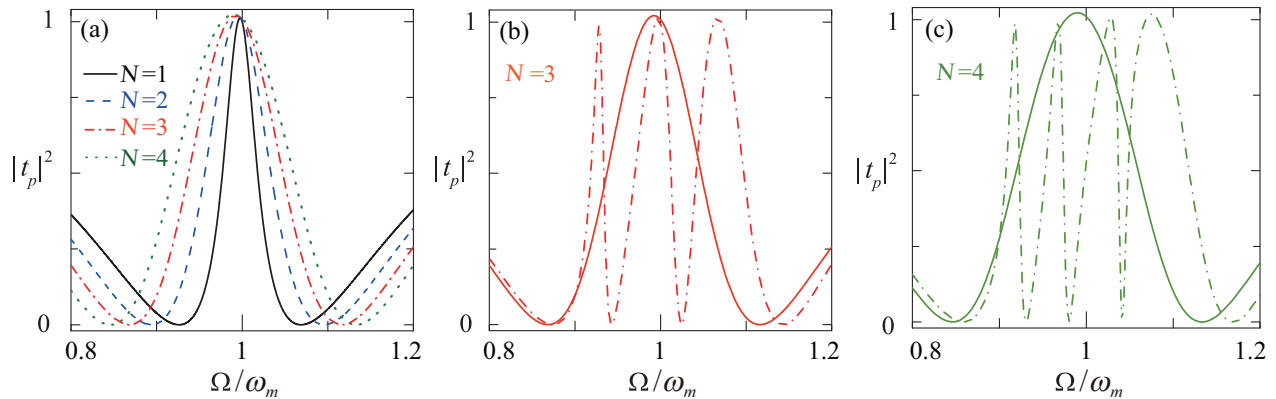}
\caption{(Color online) (a) In the presence of the dark-mode effect ($\eta=0$), the transmission rate $|t_{p}|^{2}$ of the probe light as a function of the probe-pump detuning $\Omega$ when $N=1$ (the black solid curve), $N=2$ (the blue dashed curve), $N=3$ (the red dashed-dotted curve), and $N=4$ (the dotted olive curve). The transmission rate $|t_{p}|^{2}$ of the probe light versus $\Omega$ when (b) $N=3$ and (c) $N=4$, for the dark-mode-unbreaking ($\eta=0$, the solid lines) and -breaking ($\eta/\omega_{m}=0.05$ and $\theta_{1}=\pi/2$, the dashed-dotted lines) cases. Other parameters used are $P_{L}=1.5$ mW, $g_{1}=g_{2}$, and $\Delta/\omega_{m}=1$.}\label{Darksinglefour}
\end{figure*}

In the absence of the phase-dependent phonon-phonon interaction between the adjacent mechanical modes ($H_{j}=0$), the system exists a bright mode $B_{+}=\sum_{j=1}^{N}\delta b_{j}/\sqrt{N}$ and ($N-1$) dark modes decoupled from the cavity-field mode $a$. As a result, there exists only one route of destructive interference in the $N$-degenerate-mechanical-mode optomechanical system. To see this, we plot the transmission rate of the probe light $|t_{p}|^{2}$ as a function of the probe-pump detuning $\Omega$ with the increase of the number of mechanical modes, as shown in Fig.~\ref{Darksinglefour}(a). It is obviously shown that the OMIT window can be amplified by increasing the number of mechanical modes. Physically, this can be explained by the enhanced effective mechanical decay rate of the bright mode, which is given by
\begin{eqnarray}
\label{effectdecayN}
\Gamma_{\text{effect}} &=&\gamma_{m}+N\gamma_{\text{opt}}.
\end{eqnarray}
Equation~(\ref{effectdecayN}) shows that, in comparison with the typical optomechanical system, approximately $N$ times amplification can be observed for the OMIT linewidth [see Fig.~\ref{Darksinglefour}(a)]. This study provides a route to realize the optical transmission within a large frequency bandwidth.

To control the dark-mode effect in this $N$-mechanical-mode optomechanical system, we introduce the phase-dependent phonon-phonon couplings ($H_{j}\neq0$) between the neighboring mechanical modes. We can diagonalize the Hamiltonian of these coupled mechanical modes as~\cite{Yao2011PRL,Qin2013PRA}
\begin{eqnarray}
H_{\text{mrt}} &=&\omega_{m}\sum_{j=1}^{N}\delta b_{j}^{\dagger}\delta b_{j}+\eta\sum_{j=1}^{N-1}(e^{i\theta_{j}}\delta b_{j}^{\dagger}\delta b_{j+1}+e^{-i\theta_{j}}\delta b_{j+1}^{\dagger}\delta b_{j})\nonumber\\
&=&\sum_{k=1}^{N}\Omega_{k}B_{k}^{\dagger}B_{k},
\end{eqnarray}
where $B_{k}$ describes the $k$th diagonalized mechanical mode with the resonance frequency
\begin{eqnarray}
\Omega_{k}=\omega_{m}+2\eta\cos \left(\frac{k\pi}{N+1}\right).
\end{eqnarray}
The relationship between $\delta b_{j}$ and $B_{k}$ is given by
\begin{equation}
\delta b_{j}=\bigg\{\begin{array}{c}
\frac{1}{A}\sum_{k=1}^{N}\sin \left(\frac{k\pi}{N+1}\right)B_{k}, \hspace{2.6 cm} j=1, \\
\frac{1}{A}e^{-i\sum_{\nu =1}^{j-1}\theta_{\nu }}\sum_{k=1}^{N}\sin\left(\frac{jk\pi}{N+1}\right)B_{k}, \hspace{1.3 cm} j\geq 2,
\end{array}
\end{equation}
where $A=\sqrt{(N+1)/2}$. The Hamiltonian in Eq.~(\ref{eq1iniH1Nm}) can be rewritten with these diagonalized mechanical normal modes as
\begin{eqnarray}
H_{\text{RWA}}=&\Delta\delta a^{\dagger}\delta a+\sum_{k=1}^{N}\Omega_{k}B_{k}^{\dagger}B_{k}+H_{\text{om}},
\end{eqnarray}
where the Hamiltonian $H_{\text{om}}$ with the optomechanical coupling reads as
\begin{equation}
H_{\text{om}}=\frac{G}{A}\sum_{k=1}^{N}\left[\sin\left(\frac{k\pi}{N+1}\right)+\sum_{j=2}^{N}e^{i\varphi}\sin \left(\frac{jk\pi}{N+1}\right)\right] aB_{k}^{\dagger}+\mathrm{H.c.},\label{Hom1}
\end{equation}
with
\begin{equation}
\varphi=\sum_{\nu =1}^{j-1}\theta_{\nu}.\label{phase1}
\end{equation}
We can find from Eq.~(\ref{Hom1}) that the function of these modulation phases is dominated by Eq.~(\ref{phase1}). Hence, the dark-mode-broken effect can be realized by modulating a single phase. For simplicity, below we consider $\theta_{j}=0$ for $j=2$-$(N-1)$. For an even $k$, the interaction between the cavity mode and the mechanical normal mode $B_{k}$ can be expressed as (see Appendix~\ref{appendixc})
\begin{equation}
H_{\text{ck}}=\frac{G}{A}\left[\left(1-e^{i\theta_{1}}\right) \sin \left(\frac{k\pi}{N+1}\right)\right]aB_{k}^{\dagger }+\mathrm{H.c.},\hspace{0.3 cm}k=\text{even}.\label{ck}
\end{equation}
We see from Eq.~(\ref{ck}) that when $\theta_{1}=2n\pi$, the coupling strength between the $k$th mechanical normal mode $B_{k=\text{even}}$ and the cavity mode $a$ is equal to zero. In this case, all the even normal modes are decoupled from the cavity field. Thus, the dark modes can be broken by choosing a proper parameter ($\theta_{1}\neq2n\pi$).

When switching from the dark-mode-unbreaking to -breaking cases, a single transparency window is divided into multiple transparency windows, and the number of the transparency windows depends on the number of mechanical modes [see Figs.~\ref{Darksinglefour}(b) and ~\ref{Darksinglefour}(c)]. Physically, this is the result of the destructive interferences between the weak probe field and the multiple anti-Stokes fields which are generated by the interactions of the strong coupling field and the dressed mechanical modes. These results indicate that we can steer the switch from single- to multi-channel optical communications by breaking the dark-mode effect.

\section{Conclusion\label{sec5}}

In summary, we presented a theoretical proposal to enhance and steer the OMIT, second-order sidebands, and group delay by controlling the dark-mode effect. In the three-mode loop-coupled configuration, the combination of the phase-dependent phonon-phonon interaction between the two mechanical modes with the optomechanical couplings controls the dark-mode effect via the quantum interference effect. In the presence of the dark mode, there exists only one OMIT window in the two-mechanical-mode optomechanical system, and this transmission window is nearly two-fold amplified compared to that in the typical optomechanical system. We also studied the switching from the OMIT with a single window to that of the tunable double windows by breaking the dark-mode effect. Moreover, the controllable second-order sidebands and the slow-to-fast or fast-to-slow light switching can be achieved by controlling the dark-mode effect. Finally, we extended this method to investigate the optical transmission in an $N$-mechanical-mode optomechanical system.

\begin{acknowledgments}
D.-G. L. thanks Ya-Feng Jiao, Fen Zou, and Dr. Tao Liu for valuable discussions. D.-G.L. is supported in part by Postgraduate Scientific Research Innovation Project of Hunan Province (Grant No.~CX2018B290). B.-P.H. is supported in part by NNSFC (Grant No.~11974009). J.-Q.L. is supported in part by NNSFC (Grants No.~11822501, No.~11774087, and No.~11935006), Natural Science Foundation of Hunan Province, China (Grant No.~2017JJ1021), and Hunan Science and Technology Plan Project (Grant No. 2017XK2018). F.N. is supported in part by the:
MURI Center for Dynamic Magneto-Optics via the Air Force Office of Scientific Research (AFOSR) (FA9550-14-1-0040), Army Research Office (ARO) (Grant No. Grant No. W911NF-18-1-0358),
Japan Science and Technology Agency (JST) (via the Q-LEAP program, and the CREST Grant No. JPMJCR1676), Japan Society for the Promotion of Science (JSPS) (JSPS-RFBR Grant No. 17-52-50023, and JSPS-FWO Grant No. VS.059.18N), the RIKEN-AIST Challenge Research Fund, the Foundational Questions Institute (FQXi), and the NTT PHI Laboratory.
\end{acknowledgments}

\appendix
\begin{widetext}
\section{\label{appendixa} The algebra equations and sideband parameters}
In this appendix, we present $12$ algebra equations divided into two groups and the sideband parameters used in Eqs.~(\ref{firstorder}) and ~(\ref{secondorder}).
The linear response of the probe field is described by the first group,
\begin{eqnarray}
\label{firsteq}
\varepsilon_{p}&=&[\kappa +i(\Delta -\Omega )]A_{1}^{-}+i[g_{1}\alpha(B_{1,1}^{-}+(B_{1,1}^{+})^{\ast})+g_{2}\alpha (B_{2,1}^{-}+(B_{2,1}^{+})^{\ast})],  \notag \\
0 &=&[\gamma _{1}+i(\omega _{1}-\Omega )]B_{1,1}^{-}+i[g_{1}(\alpha ^{\ast}A_{1}^{-}+\alpha (A_{1}^{+})^{\ast})+\eta e^{i\theta }B_{2,1}^{-}],  \notag \\
0 &=&[\gamma_{2}+i(\omega_{2}-\Omega)]B_{2,1}^{-}+i[g_{2}(\alpha ^{\ast}A_{1}^{-}+\alpha (A_{1}^{+})^{\ast})+\eta e^{-i\theta}B_{1,1}^{-}],  \notag \\
0 &=&[\kappa -i(\Delta +\Omega)](A_{1}^{+})^{\ast }-i[g_{1}\alpha ^{\ast }( B_{1,1}^{-}+(B_{1,1}^{+})^{\ast})+g_{2}\alpha ^{\ast}( B_{2,1}^{-}+(B_{2,1}^{+})^{\ast})],  \notag \\
0 &=&[\gamma _{1}-i(\omega _{1}+\Omega)](B_{1,1}^{+})^{\ast }-i[g_{1}(\alpha ^{\ast}A_{1}^{-}+\alpha(A_{1}^{+})^{\ast})+\eta e^{-i\theta }(B_{2,1}^{+})^{\ast }],  \notag \\
0 &=&[\gamma _{2}-i(\omega _{2}+\Omega)](B_{2,1}^{+})^{\ast }-i[g_{2}(\alpha ^{\ast}A_{1}^{-}+\alpha(A_{1}^{+})^{\ast})+\eta e^{i\theta }\left( B_{1,1}^{+}\right)^{\ast }],
\end{eqnarray}
while the second-order sideband is exhibited by the second group,
\begin{eqnarray}
\label{secondeq}
0 &=&[\kappa +i(\Delta -2\Omega )]A_{2}^{-}+i\{g_{1}[\alpha B_{1,2}^{-}+\alpha (B_{1,2}^{+})^{\ast}+A_{1}^{-}(B_{1,1}^{-}+(B_{1,1}^{+})^{\ast })]+g_{2}[\alpha(B_{2,2}^{-}+(B_{2,2}^{+})^{\ast})+A_{1}^{-}(B_{2,1}^{-}+(B_{2,1}^{+})^{\ast })]\},  \notag \\
0 &=&[\gamma _{1}+i(\omega _{1}-2\Omega )]B_{1,2}^{-}+i[g_{1}(\alpha ^{\ast}A_{2}^{-}+\alpha (A_{2}^{+})^{\ast}+(A_{1}^{+})^{\ast }A_{1}^{-})+\eta e^{i\theta }B_{2,2}^{-}],  \notag \\
0 &=&[\gamma _{2}+i(\omega _{2}-2\Omega )]B_{2,2}^{-}+i[g_{2}(\alpha ^{\ast}A_{2}^{-}+\alpha( A_{2}^{+})^{\ast}+(A_{1}^{+})^{\ast }A_{1}^{-})+\eta e^{-i\theta }B_{1,2}^{-}],  \notag \\
0 &=&(\kappa-i\Delta -2i\Omega)( A_{2}^{+})^{\ast}-i\{g_{1}[\alpha ^{\ast }B_{1,2}^{-}+\alpha ^{\ast }(B_{1,2}^{+})^{\ast}+(A_{1}^{+})^{\ast }B_{1,1}^{-}+(A_{1}^{+})^{\ast}(B_{1,1}^{+})^{\ast}]+g_{2}[\alpha^{\ast}B_{2,2}^{-}+\alpha ^{\ast}(B_{2,2}^{+})^{\ast} \notag \\
&&+(A_{1}^{+})^{\ast }B_{2,1}^{-}+(A_{1}^{+})^{\ast}(B_{2,1}^{+})^{\ast}]\},  \notag \\
0 &=&(\gamma_{1}-i\omega_{1}-2i\Omega) (B_{1,2}^{+})^{\ast }-i[g_{1}(\alpha ^{\ast }A_{2}^{-}+\alpha(A_{2}^{+})^{\ast}+(A_{1}^{+})^{\ast }A_{1}^{-})+\eta e^{-i\theta}(B_{2,2}^{+})^{\ast}],  \notag \\
0 &=&(\gamma_{2}-i\omega _{2}-2i\Omega)(B_{2,2}^{+})^{\ast }-i[g_{2}(\alpha^{\ast}A_{2}^{-}+\alpha(A_{2}^{+})^{\ast}+(A_{1}^{+}) ^{\ast }A_{1}^{-})+\eta e^{i\theta }(B_{1,2}^{+})^{\ast}].
\end{eqnarray}

The other parameters used in Eqs.~(\ref{firstorder}) and ~(\ref{secondorder}) are
\begin{eqnarray}
\label{paremeters2}
B_{1,1}^{-} &=&\frac{g_{1}\alpha ^{\ast }V_{1}\left[ \gamma _{2}+i\left(
\omega _{2}-\Omega \right) \right] -ig_{2}\alpha ^{\ast }V_{1}\eta
e^{i\theta }}{\left( i\kappa +\Delta +\Omega \right) \left\{ \left(
4\left\vert \alpha \right\vert ^{2}\Delta \left(
g_{2}^{2}T_{3,1}^{(1)}+g_{1}^{2}T_{3,2}^{(1)}\right) -T_{2}^{(1)}\left[
\Delta ^{2}+\left( \kappa -i\Omega \right) ^{2}\right] \right)
+8g_{1}g_{2}\eta \left\vert \alpha \right\vert ^{2}T_{1}^{(1)}\Delta \cos
\theta \right\} }\varepsilon_{p} ,\\
B_{2,1}^{-} &=&\frac{g_{2}\alpha ^{\ast }V_{1}\left[ \gamma _{1}+i\left(
\omega _{1}-\Omega \right) \right] -ig_{1}\alpha ^{\ast }V_{1}\eta
e^{-i\theta }}{\left( i\kappa +\Delta +\Omega \right) \left\{ \left(
4\left\vert \alpha \right\vert ^{2}\Delta \left(
g_{2}^{2}T_{3,1}^{(1)}+g_{1}^{2}T_{3,2}^{(1)}\right) -T_{2}^{(1)}\left[
\Delta ^{2}+\left( \kappa -i\Omega \right) ^{2}\right] \right)
+8g_{1}g_{2}\eta \left\vert \alpha \right\vert ^{2}T_{1}^{(1)}\Delta \cos
\theta \right\} }\varepsilon_{p}, \\
\left( B_{1,1}^{+}\right) ^{\ast } &=&\frac{-iV_{2}\left[ g_{1}\alpha ^{\ast
}\left( i\gamma _{2}+\omega _{2}+\Omega \right) -g_{2}\alpha ^{\ast }\eta
e^{-i\theta }\right] }{\left( i\kappa +\Delta +\Omega \right) \left\{ \left(
4\left\vert \alpha \right\vert ^{2}\Delta \left(
g_{2}^{2}T_{3,1}^{(1)}+g_{1}^{2}T_{3,2}^{(1)}\right) -T_{2}^{(1)}\left[
\Delta ^{2}+\left( \kappa -i\Omega \right) ^{2}\right] \right)
+8g_{1}g_{2}\eta \left\vert \alpha \right\vert ^{2}T_{1}^{(1)}\Delta \cos
\theta \right\} }\varepsilon_{p}, \\
\left( B_{2,1}^{+}\right) ^{\ast } &=&\frac{-e^{i\theta }g_{1}\alpha ^{\ast
}\eta V_{3}+g_{2}\alpha ^{\ast }V_{3}\left( i\gamma _{1}+\omega _{1}+\Omega
\right) }{-\left\{ \left( 4\left\vert \alpha \right\vert ^{2}\Delta \left(
g_{2}^{2}T_{3,1}^{(1)}+g_{1}^{2}T_{3,2}^{(1)}\right) -T_{2}^{(1)}\left[
\Delta ^{2}+\left( \kappa -i\Omega \right) ^{2}\right] \right)
+8g_{1}g_{2}\eta \left\vert \alpha \right\vert ^{2}T_{1}^{(1)}\Delta \cos
\theta \right\} }\varepsilon_{p},
\end{eqnarray}
where
\begin{eqnarray}
V_{1} &=&[-\eta ^{2}+(i\gamma _{1}+\omega _{1}+\Omega)(i\gamma _{2}+\omega_{2}+\Omega )][\kappa -i(\Delta +\Omega )]^{2} ,  \\
V_{2} &=&[\eta ^{2}+(-i\gamma _{1}+\omega _{1}-\Omega)(i\gamma _{2}-\omega_{2}+\Omega )][\kappa -i(\Delta +\Omega )]^{2} ,  \\
V_{3} &=&[-\eta ^{2}+(i\gamma _{1}-\omega _{1}+\Omega)(i\gamma _{2}-\omega_{2}+\Omega )][\kappa -i(\Delta +\Omega )], \\
T^{(1)}_{1} &=&-\omega_{1}\omega _{2}+\eta ^{2}+(\gamma _{1}-i\Omega)(\gamma _{2}-i\Omega),  \\
T^{(1)}_{2} &=&[\eta ^{2}+(-i\gamma _{1}+\omega_{1}-\Omega)(i\gamma _{2}-\omega_{2}+\Omega)][\eta ^{2}+(\gamma _{1}-i(\omega _{1}+\Omega))(\gamma _{2}-i(\omega _{2}+\Omega))],  \\
T^{(1)}_{3,1} &=&(\gamma_{1}^{2}+\omega_{1}^{2}-\Omega^{2}-2i\gamma_{1}\Omega)\omega_{2}-\omega_{1}\eta^{2} ,  \\
T^{(1)}_{3,2} &=&(\gamma_{2}^{2}+\omega_{2}^{2}-\Omega^{2}-2i\gamma_{2}\Omega)\omega_{1}-\omega_{2}\eta^{2},
\end{eqnarray}
and $T^{(2)}_{1}$, $T^{(2)}_{2}$, $T^{(2)}_{3,1}$, and $T^{(2)}_{3,2}$ can be obtained with replacing the $\Omega$ as $2\Omega$ in the $T^{(1)}_{1}$, $T^{(1)}_{2}$, $T^{(1)}_{3,1}$ , and $T^{(1)}_{3,2}$.
\end{widetext}

\section{\label{appendixb}Derivation of the effective mechanical decay rate of the bright mode}
In this appendix, we derive the effective mechanical decay rate of the bright mode in the two-mechanical-mode optomechanical system. We consider the case
where the phase-dependent phonon-exchange interaction and the probe field are absent (i.e., $\eta=0$ and $\varepsilon_{p}=0$). Based on Eq.~(\ref{Langevineqorig0}), the linearized Langevin equations for quantum fluctuations are given by
\begin{subequations}
\label{lineLangevineq}
\begin{align}
\delta\dot{a}=&-(\kappa +i\Delta) \delta a-i\sum_{l=1,2}G_{l}(\delta b_{l}+ \delta b_{l}^{\dagger})+\sqrt{2\kappa}a_{\text{in}},\\
\delta\dot{b}_{l=1,2}=&-iG_{l}^{\ast}\delta a-(\gamma_{l}+i\omega_{l}) \delta b_{l}-iG_{l}\delta a^{\dagger}+\sqrt{2\gamma_{l}}b_{l,\text{in}},
\end{align}
\end{subequations}
where $\Delta =\Delta _{c}+\sum_{l=1,2}g_{l}(\beta_{l}+\beta_{l}^{\ast})$ is the normalized driving detuning, and $G_{l}=g_{l}|\alpha|$ denotes the linearized optomechanical coupling strength between the cavity-field mode and the $l$th mechanical mode.

Below, we consider the case where the linearized optomechanical coupling strengths $G_{1,2}$ are real and the system works in the parameter regime $\omega_{1,2}\gg \kappa\gg G_{1,2}\gg \gamma_{1,2}$. In this case, the cavity-field mode can be eliminated adiabatically, and then the solution of the cavity field fluctuation operator $\delta a(t)$ at the time scale $t\gg 1/\kappa$ can be obtained as
\begin{eqnarray}
\delta a(t)&\approx &-\frac{iG_{1}}{\kappa +i(\Delta +\omega_{1})}\delta b_{1}^{\dagger}(t)-\frac{iG_{1}}{\kappa +i(\Delta-\omega_{1})}\delta b_{1}(t)\nonumber \\
&&-\frac{iG_{2}}{\kappa +i(\Delta+\omega_{2})}\delta b_{2}^{\dagger}(t)-\frac{iG_{2}}{\kappa +i(\Delta -\omega_{2})}\delta b_{2}(t)\nonumber\\
&&+F_{a,\text{in}}(t),\label{dletaaadiaelim}
\end{eqnarray}
where we introduce the noise operator
\begin{equation}
F_{a,\text{in}}(t)=\sqrt{2\kappa }e^{-(\kappa +i\Delta)t}\int_{0}^{t}a_{\text{in}}(s)e^{(\kappa +i\Delta)s}ds.
\end{equation}
Substitution of Eq.~(\ref{dletaaadiaelim}) into Eqs.~(\ref{lineLangevineq}b) and~(\ref{lineLangevineq}c) leads to the equations of motion
\begin{eqnarray}
\label{redeqofb1b2}
\delta\dot{b}_{1}(t)&=&-(\Gamma_{1}+i\Omega_{1})\delta b_{1}(t)+\xi_{1}\delta b_{2}(t)-iG_{1}F_{a,\text{in}}(t)\nonumber \\
&&-iG_{1}F_{a,\text{in}}^{\dagger}(t)+\sqrt{2\gamma_{1}}b_{1,\text{in}}(t),\nonumber\\
\delta\dot{b}_{2}(t)&=&\xi_{2}\delta b_{1}(t)-(\Gamma_{2}+i\Omega_{2}) \delta b_{2}(t)-iG_{2}F_{a,\text{in}}(t)\nonumber \\
&&-iG_{2}F_{a,\text{in}}^{\dagger}(t)+\sqrt{2\gamma_{2}}b_{2,\text{in}}(t),
\end{eqnarray}
where we introduce the parameters
\begin{eqnarray}
\xi_{1}&=&\frac{G_{1}G_{2}[\kappa +i(\Delta +\omega _{2})]}{\kappa^{2}+(\Delta +\omega _{2})^{2}}-\frac{G_{1}G_{2}[\kappa -i(\Delta -\omega_{2})]}{\kappa ^{2}+(\Delta -\omega _{2})^{2}}, \nonumber\\
\xi_{2}&=&\frac{G_{1}G_{2}[\kappa +i(\Delta +\omega _{1})]}{\kappa^{2}+(\Delta +\omega _{1})^{2}}-\frac{G_{1}G_{2}[\kappa -i(\Delta -\omega_{1})]}{\kappa ^{2}+(\Delta -\omega _{1})^{2}},
\end{eqnarray}
and
\begin{eqnarray}
\Gamma_{l} =\gamma_{l}+\gamma_{l,\text{opt}}, \hspace{0.5 cm}
\Omega_{l}=\omega_{l}-\omega_{l,\text{opt}},
\end{eqnarray}
with
\begin{eqnarray}
\gamma_{l,\text{opt}}&=&\frac{G_{l}^{2}\kappa}{\kappa^{2}+(\Delta-\omega_{l})^{2}}-\frac{G_{l}^{2}\kappa}{\kappa^{2}+(\Delta+\omega _{l})^{2}},\nonumber\\
\omega_{l,\text{opt}}&=&\frac{G_{l}^{2}(\Delta+\omega_{l})}{\kappa^{2}+(\Delta +\omega_{l})^{2}}+\frac{G_{l}^{2}(\Delta -\omega_{l})}{\kappa^{2}+(\Delta-\omega_{l})^{2}},\hspace{0.3 cm} l=1,2.
\end{eqnarray}
Under the parameter condition $\omega_{1,2}\gg\kappa\gg G_{1,2}$ and at resonance $\Delta=\omega_{1}=\omega_{2}$, we have
\begin{eqnarray}
\xi_{1}\approx&-[G_{1}G_{2}/\kappa -i(G_{1}G_{2}/2\omega_{2})],\nonumber\\
\xi_{2}\approx&-[G_{1}G_{2}/\kappa -i(G_{1}G_{2}/2\omega_{1})],
\end{eqnarray}
and
\begin{eqnarray}
\gamma_{l,\text{opt}}\approx G_{l}^{2}/\kappa,\hspace{0.3 cm}\omega_{l,\text{opt}}\approx G_{l}^{2}/(2\omega_{l}),\hspace{0.5 cm} l=1,2.
\end{eqnarray}
Here we assume that the two mechanical modes have the identical resonance frequencies ($\omega_{l}=\omega_{m}$), decay rates ($\gamma_{l}=\gamma_{m}$), and optomechanical couplings ($G_{l}=G$). Thus, we can obtain the optomechanically induced mechanical decay rate $\gamma_{l,\text{opt}}=\gamma_{\text{opt}}$ and the resonant frequency $\omega_{l,\text{opt}}=\omega_{\text{opt}}$. Then, the equations of motion for the bright mode $B_{+}$ can be written as
\begin{eqnarray}
\dot{B}_{+} &=&\frac{\dot{b}_{1}+\dot{b}_{2}}{\sqrt{2}}=-(\Gamma _{\text{eff}}+i\Omega _{\text{eff}})B_{+}  \notag \\
&&+2iG_{1}F_{\text{in}}+\sqrt{2\gamma _{m}}B_{+,\text{in}},
\end{eqnarray}
where
\begin{subequations}
\begin{align}
\Gamma_{\text{eff}}=&\gamma_{m}+2\gamma_{\text{opt}}, \\
\Omega_{\text{eff}}=&\omega_{m}-2\omega_{\text{opt}}, \\
B_{+,\text{in}}=&[b_{1,\text{in}}(t) +b_{2,\text{in}}(t)]/\sqrt{2}, \\
F_{\text{in}}=&[F_{a,\text{in}}(t) +F_{a,\text{in}}^{\dagger}(t)]/\sqrt{2}.
\end{align}
\end{subequations}
Here, $\Gamma _{\text{eff}}$ and $\Omega _{\text{eff}}$ are, respectively, the effective mechanical decay rate and resonant frequency of the bright mode $B_{+}$.

\section{\label{appendixc}Derivation of Eq.~(\ref{ck})}

In this appendix, we show a detailed derivation of Eq.~(\ref{ck}). Based on Eqs.~(\ref{Hom1}) and ~(\ref{phase1}), for $N\geq3$, we obtain the effective coupling coefficient between the cavity $a$ and the mode $B_{k}$ as
\begin{eqnarray}
&&\frac{G}{A}\left[\sin \left(\frac{k\pi}{N+1}\right)+\sum_{j=2}^{N}e^{i\sum_{\nu =1}^{j-1}\theta _{\nu }}\sin \left(\frac{jk\pi }{N+1}\right)\right] \nonumber\\
&=&\frac{G}{A}\bigg\{\Big[\sin \left(\frac{1}{N+1}k\pi \right)+e^{i\sum_{\nu=1}^{N-1}\theta _{\nu }}\sin \left(\frac{N}{N+1}k\pi\right)\Big]\nonumber\\
&&+\Big[e^{i\theta_{1}}\sin \left(\frac{2}{N+1}k\pi\right) +e^{i\sum_{\nu =1}^{N-2}\theta _{\nu }}\sin \left(\frac{N-1}{N+1}k\pi\right)\Big]\nonumber\\
&&+\Big[e^{i(\theta_{1}+\theta _{2})}\sin \left(\frac{3}{N+1}k\pi\right)+e^{i\sum_{\nu =1}^{N-3}\theta_{\nu }}\sin \left(\frac{N-2}{N+1}k\pi\right)\Big]\nonumber\\
&&+\cdots\bigg\}.
\end{eqnarray}
Below, we consider two cases corresponding to odd and even numbers $N$, respectively.

(i) We firstly consider the case where $N$ is an odd number and $ \theta_{j}=0$ for $j=2$-$(N-1)$, and the coefficient becomes
\begin{eqnarray}
&&\frac{G}{A}\left[\sin \left(\frac{k\pi }{N+1}\right)+\sum_{j=2}^{N}e^{i\sum_{\nu =1}^{j-1}\theta _{\nu }}\sin \left(\frac{jk\pi}{N+1}\right)\right] \nonumber\\
&=&\frac{G}{A}\bigg\{\left[\sin \left(\frac{k\pi}{N+1}\right)+e^{i\theta _{1}}\sin\left(\frac{Nk\pi}{N+1}\right)\right]\nonumber\\
&&+e^{i\theta _{1}}\left[\sin \left(\frac{2k\pi}{N+1}\right)+\sin \left(\frac{N-1}{N+1}k\pi\right) \right] \nonumber\\
&&+e^{i\theta_{1}}\left[\sin \left(\frac{3k\pi}{N+1}\right)+\sin \left(\frac{N-2}{N+1}k\pi\right)\right]\nonumber\\
&&+\cdots+e^{i\theta _{1}}\sin \left(\frac{k\pi }{2}\right)\bigg\}.\label{odd}
\end{eqnarray}
When $k$ is an odd number, we have
\begin{eqnarray}
&&\frac{G}{A}\left[\sin \left(\frac{k\pi }{N+1}\right)+\sum_{j=2}^{N}e^{i\sum_{\nu =1}^{j-1}\theta_{\nu }}\sin \left(\frac{jk\pi }{N+1}\right)\right] \nonumber\\
&=&\frac{G}{A}\bigg[(1+e^{i\theta _{1}}) \sin \left(\frac{k\pi}{N+1}\right)+2e^{i\theta_{1}}\sin \left(\frac{2k\pi }{N+1}\right)\nonumber\\
&&+2e^{i\theta_{1}}\sin \left(\frac{3k\pi}{N+1}\right)+\cdots+e^{i\theta _{1}}\sin\left (\frac{k\pi}{2}\right)\bigg];\label{odd1}
\end{eqnarray}
When $k$ is an even number, we obtain
\begin{eqnarray}
&&\frac{G}{A}\left[\sin \left(\frac{k\pi}{N+1}\right)+\sum_{j=2}^{N}e^{i\sum_{\nu =1}^{j-1}\theta_{\nu }}\sin \left(\frac{jk\pi }{N+1}\right)\right]\nonumber\\
&=&\frac{G}{A}\left( 1-e^{i\theta _{1}}\right)\sin\left(\frac{k\pi}{N+1}\right).\label{odd2}
\end{eqnarray}
(ii) Then, we consider the case where $N$ is an even number and $ \theta_{j}=0$ for $j=2$-$(N-1)$, and the coefficient can be simplified as
\begin{eqnarray}
&&\frac{G}{A}\left[\sin \left(\frac{k\pi}{N+1}\right)+\sum_{j=2}^{N}e^{i\sum_{\nu =1}^{j-1}\theta _{\nu }}\sin \left(\frac{jk\pi }{N+1}\right)\right] \nonumber\\
&=&\frac{G}{A}\bigg\{\left[\sin \left(\frac{k\pi }{N+1}\right)+e^{i\theta_{1}}\sin\left(\frac{N}{N+1}k\pi\right)\right]\nonumber\\
&&+e^{i\theta _{1}}\left[\sin \left(\frac{2k\pi }{N+1}\right)+\sin \left(\frac{N-1}{N+1}k\pi\right)\right] \nonumber\\
&&+e^{i\theta_{1}}\left[\sin \left(\frac{3k\pi }{N+1}\right)+\sin \left(\frac{N-2}{N+1}k\pi\right)\right]+\cdots\bigg\}.\label{even}
\end{eqnarray}
If $k$ is an odd number, we obtain
\begin{eqnarray}
&&\frac{G}{A}\left[\sin \left(\frac{k\pi }{N+1}\right)+\sum_{j=2}^{N}e^{i\sum_{\nu =1}^{j-1}\theta _{\nu }}\sin \left(\frac{jk\pi }{N+1}\right)\right] \nonumber\\
&=&\frac{G}{A}\bigg[(1+e^{i\theta _{1}})\sin \left(\frac{k\pi }{N+1}\right)+2e^{i\theta _{1}}\sin \left(\frac{2k\pi }{N+1}\right)\nonumber\\
&&+2e^{i\theta_{1}}\sin \left(\frac{3k\pi}{N+1}\right)+\cdots\bigg];\label{even1}
\end{eqnarray}
If $k$ is an even number, we have
\begin{eqnarray}
&&\frac{G}{A}\left[\sin\left(\frac{k\pi}{N+1}\right)+\sum_{j=2}^{N}e^{i\sum_{\nu =1}^{j-1}\theta _{\nu }}\sin \left(\frac{jk\pi}{N+1}\right)\right]\nonumber\\
&=&\frac{G}{A}\left(1-e^{i\theta_{1}}\right)\sin \left(\frac{k\pi}{N+1}\right).\label{even2}
\end{eqnarray}

According to Eqs.~(\ref{odd}-\ref{even2}), we can summarize that for an even number $k$, the optomechanical interaction between the mechanical mode $B_{k=\text{even}}$ and the cavity mode $a$ is described by Eq.~(\ref{ck}).

\end{document}